\documentclass[10pt,conference]{IEEEtran}
\usepackage{amsmath, amssymb, bm, cite, epsfig, psfrag}
\usepackage{graphicx}
\usepackage[top    = 0.75in,
bottom = 1in,
left   = 0.625in,
right  = 0.625in]{geometry}
\usepackage{dblfloatfix}
\usepackage{array}
\usepackage{textcomp}
\usepackage{float}
\usepackage{longtable}
\usepackage{supertabular,booktabs}
\usepackage{multirow}
\usepackage[usenames,dvipsnames]{xcolor}

\usepackage{etoolbox}
\usepackage{pbox}
\usepackage{fixltx2e}
\usepackage{romannum}
\usepackage{filecontents}
\usepackage{enumerate}
\graphicspath{ {figures/} }
\usepackage{subfigure}
\usepackage{colortbl}
\usepackage{fancyhdr}

\usepackage{bm}
\newtoggle{conference}
\togglefalse{conference} %
\graphicspath{{figures/}}

\fancypagestyle{firststyle}{
	\fancyhf{}
	\fancyhead[L]{O. Kanhere and T. S. Rappaport, ``Outdoor sub-THz Position Location and Tracking using Field Measurements at 142 GHz,'' \textit{in 2021 IEEE International Conference on Communications (ICC), June 2021, pp. 1--6.}  }   %

}
\usepackage{etoolbox}
\makeatletter
\patchcmd{\@makecaption}
{\scshape}
{}
{}
{}
\makeatletter
\patchcmd{\@makecaption}
{\\}
{.\ }
{}
{}
\makeatother
\IEEEoverridecommandlockouts
\begin{document}
\title{Outdoor sub-THz Position Location and Tracking using Field Measurements at 142 GHz}
\author{\IEEEauthorblockN{Ojas Kanhere, Theodore S. Rappaport \thanks{This work is supported by the NYU WIRELESS Industrial Affiliates Program and National Science Foundation (NSF) Grants: 1909206 and 2037845.}}
	
\IEEEauthorblockA{	\small NYU WIRELESS\\
					 Tandon School of Engineering, New York University\\
					Brooklyn, NY 11201\\
					\{ojask, tsr\}@nyu.edu}}

\maketitle
\thispagestyle{firststyle}

\begin{abstract}

 Future sub-THz cellular deployments may be utilized to complement the coverage of the global positioning system (GPS) and provide centimeter-level accuracy. In this work, we use measurement data at 142 GHz to test a map-based position location algorithm in an outdoor urban microcell (UMi) environment. We utilize an extended Kalman filter (EKF) to track the position of the user equipment (UE) along a rectangular track, with the transmitter-receiver separation distances varying from 24.3 m to 52.8 m. The position and velocity of the UE are tracked by the EKF, with measurements of the angle of arrival and time of flight information obtained along an outdoor track, to provide a mean accuracy of 24.8 cm at 142 GHz, over 34 UE locations, using a single base station in line-of-sight and non-line-of-sight. 

\end{abstract}
    

\section{Introduction}\label{Introduction}

Accurate knowledge of the position of mobile devices is crucial for a variety of outdoor applications, such as navigation in unknown environments and tracking the location of commercial vehicular fleets. The global positioning system (GPS) can achieve centimeter-level accuracy with real-time kinematics (RTK), where the user equipment (UE) measures the carrier-phase differential between satellites\cite{Kozlov_1998}. However, GPS coverage is inadequate in urban canyons with high-rise buildings, in underground parking areas, and indoors, where the GPS reference signal is blocked by building walls. 

Current millimeter wave (mmWave) cellular networks and future sub-THz cellular networks are prime candidates to complement GPS for position location \cite{Rappaport19a}. Wide bandwidth is available at mmWave frequencies, with channel bandwidths up to 400 MHz configurable in the frequency range 2 (FR2) of the fifth generation of mobile technologies (5G) \cite{3GPP.38.101.2}. Phased arrays with hundreds of antenna elements and narrow half power beamwidths (HPBW) are commercially available at mmWave frequencies. The sixth generation of mobile technologies (6G) will see a move to frequencies above 100 GHz (sub-THz frequencies) where channel allocations spanning several GHz are feasible and where precise angle of arrival (AoA) and time of flight (ToF) resolution are supported \cite{Rappaport19a, Xing21a, Xing21b,Kanhere_2021}. 

A popular method for outdoor position location is fingerprinting, wherein channel parameters (such as received signal strength or channel state information) are first measured and recorded for reference points with known positions. Once a database of the a priori measured channel parameters is created, a prediction model is used to determine the unknown position of a UE based on the previously measured channel parameters. The authors of \cite{Gante_2020} used temporal convolutional networks to predict the unknown position of a UE. The UE measured and stored the power delay profile (PDP) of pulsed waveforms transmitted by a BS over a fixed beam codebook to create a \textit{beamfored fingerprint} at the UE. The beamformed fingerprint was sent back from the UE to the BS, for UE position location using a temporal convolutional network. An average position location error of 1.78 m was achieved in simulations at 28 GHz over a 400 $ \times $ 400 m area in \cite{Gante_2020}. Real-world outdoor field measurements at 28 GHz were used in \cite{H_Sun_2020} to analyze the performance of fingerprinting-based outdoor localization. Using a 500 MHz wideband channel frequency response as the signal characteristic, the fingerprinting algorithm could differentiate between 8 UE locations spaced 1 meter apart. 

Unlike fingerprinting algorithms that are data-driven, geometry-based position location algorithms that exploit multipath components (MPCs) have also been investigated for position location. The outdoor mmWave position location accuracy of a MPC-based position location algorithm was demonstrated via simulations \cite{Ruble_2018}, where the authors utilized the ToF, AoA, and angle of departure (AoD) of the MPCs to estimate the position of the UE in line-of-sight (LOS) and non-line-of-sight (NLOS). Two outdoor urban scenarios were considered: an urban canyon and an urban corner. The authors assumed NLOS MPCs measured by the UE arrived after one bounce \cite{Ruble_2018}. A median accuracy of 50 cm was achieved over an area of 50  $ \times $ 15 m. 

Map assisted positioning with angle and time (MAP-AT) is a map-based positioning approach that fuses angular and temporal information with a map of the environment to provide centimeter-level position location in LOS and NLOS environments. The MAP-AT approach makes no assumption on the number of reflections encountered by the NLOS MPC. The performance of MAP-AT was examined with real-world indoor field measurements in \cite{Kanhere20a}. An excellent mean accuracy of 5.7 cm at 28 GHz and 6.3 cm at 140 GHz was achieved over distances ranging from 4.2 m to 32.3 m. 

In addition to utilizing channel measurements to determine the current position of the UE, tracking the path of a mobile UE has been well-investigated. In \cite{Koivisto_2016}, the authors use the unscented Kalman filter (UKF) and extended Kalman filter (EKF) to simultaneously determine the position of a moving vehicle and for network synchronization. A root mean squared error of 0.76 m was achieved over the vehicular trajectory. In \cite{Rastorgueva_2018}, the authors used an EKF to track the position of a UE moving along a 100 m long trajectory by estimating the angle of departure, to achieve sub-meter positioning accuracy 90\% of the time at an operating frequency of 39 GHz.

In this paper, we extend the work in \cite{Kanhere20a} by evaluating the performance of MAP-AT in an outdoor environment and demonstrate that the EKF tracking algorithm works well and could be implemented at a BS transmitting in the sub-THz frequency band of 140 GHz, for UE tracking.

The remainder of this paper is organized as follows. The MAP-AT method and position location algorithm is described in Section \ref{sec:MAP}, along with the design of an EKF tracking algorithm. The performance of MAP-AT localization and tracking on real-world outdoor measurement data at 140 GHz is evaluated in Section \ref{sec:results}. Concluding remarks and directions for future work are provided in Section \ref{sec:conclusion}.

\section{Position Location and Tracking with MAP-AT}\label{sec:MAP}

MAP-AT is a position location technique that fuses the AoA and ToF of MPCs and associates them with a map of the environment (that is pre-generated or generated on-the-fly \cite{H_Wang_2005,Kanhere19a}). The AoA of an uplink reference signal may be measured by the phased array at the base station (BS). Note that an alternate implementation of MAP-AT may utilize downlink reference signals by measuring the AoD of the MPCs at the BS, as described in \cite{Kanhere19a}. The ToF may be measured using the round trip time of the reference signal in order to avoid synchronization issues.

\subsection{Position Location with MAP-AT }
MAP-AT makes no assumptions on the number of reflections suffered by each MPC since strong reflections with multiple bounces are possible at mmWave and sub-THz frequencies \cite{Ju20}. MAP-AT is capable of determining the location of the UE in LOS and NLOS, provided at least two MPCs arrive at the UE.

\begin{figure}
	\centering
	\includegraphics[width=0.30\textwidth]{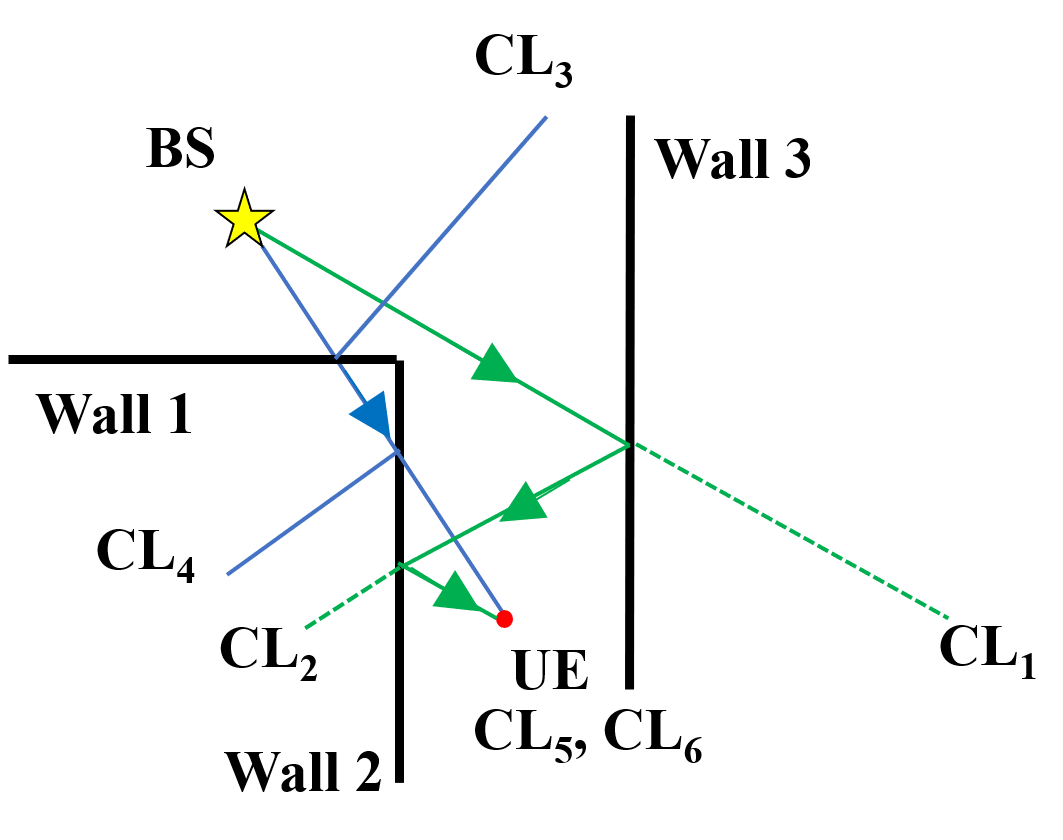}
	\caption{ Two NLOS MPCs arrive at the UE shown above - one in green and one in blue. Two candidate locations $ (\text{ CL}_5, \text{ CL}_6) $ are at the actual user location, while the other candidate locations are dispersed \cite{Kanhere19a,Kanhere20a}. }
	\label{fig:candidates}
\end{figure}

Using a site-specific map of the environment, MAP-AT back-traces the path of each MPC received at the BS. On encountering an obstruction, MAP-AT traces the ray that would have been reflected by the obstruction as well as the ray that would have penetrated through the obstruction. In this manner, MAP-AT maps out all possible UE locations (henceforth referred to as \textit{candidate locations}) based on the AoA and ToF of each MPC, measured at the BS. When two or more MPCs are back-traced, since a majority of the candidate locations coincide with the true UE location \cite{Kanhere19a,Kanhere20a}, MAP-AT can easily determine the position of the UE. Fig. \ref{fig:candidates} depicts a UE in an outdoor environment, which receives two MPCs from the BS located around the building corner. One MPC arrives at the UE after suffering two reflections while the other MPC arrives after penetrating through the building. All the candidate locations are illustrated in the figure. Note that two candidate locations coincide with or are clustered around the true UE location on the map while the other candidate locations are dispersed.

\subsection{Location tracking with the EKF} \label{sec:EKF}
Prior MAP-AT work \cite{Kanhere19a,Kanhere20a} focused on determining the UE location at a single time instant. We now extend MAP-AT for user tracking, i.e. following the position location of a mobile UE. Tracking the user location reduces the effect of sudden error spikes in AoA or ToF measurements, and provides an estimate of the UE location over motion even when the UE experiences a temporary outage. 

An EKF may be used to track the position of the UE. The EKF acts like a low pass filter, smoothening the error in the position along the track of a user. In this work, we assume a constant velocity model for the EKF, with known system input \cite{Leibowicz_2000,Yang_2018}. Varying velocity can be accommodated by estimating the UE velocity by measuring the Doppler shift of UE transmissions over the air or with onboard gyroscopes, accelerometers, and other sensors in the UE. 

The position and velocity of the UE at time instant $ \mathbf{k} $ are represented by state vector $\mathbf{x_k}= [x_k,v_{xk},y_k,v_{yk} ]^T$.  The evolution of the state of a UE with time (the relationship betweem $\mathbf{{x_{k-1}}}$ and $\mathbf{ x_k }$) is described by state dynamics. The state dynamics may be described mathematically as \cite{Li_2003}:
\begin{align}
	\mathbf{x_k} &= \mathbf{F_{k-1} x_{k-1}} + \mathbf{u_{k-1}} + \mathbf{w_{k-1}}.
\end{align}
 $ \mathbf{F_k} $ is the state transition matrix which represents the linear relationship between $\mathbf{{x_{k-1}}}$ and $\mathbf{ x_k }$. At time instants where the velocity of the UE did not change, $ \mathbf{u_{k-1}} $, the control input is equal to zero and $ \mathbf{F_k} $ is given by \cite{Li_2003,Saho_2015}:
\begin{align}
	\mathbf{F_k}  &= \begin{bmatrix} 
	1 & T & 0 &0  \\
	0 & 1 &0 &0\\
	0 & 0 & 1 &T  \\
	0 & 0 & 0 &1 
	\end{bmatrix}. 	
\end{align}
with $ T $ equal to the sampling period in seconds. 

For a UE moving along a rectangular track at a constant speed, the velocity of the UE changes four times (at the four corners of the rectangular track) due to the change in direction of motion. At the four corners of the rectangular track,  $ \mathbf{F_k} $ is equal to the 4 $ \times $ 4 zero matrix, while $\mathbf{u_{k-1}}$ specifies the state of the UE. At the four corners of the rectangle, $\mathbf{u_{k-1}}$ is equal to the state (position and velocity) of the UE after the turn, assuming the UE continues to move along the rectangular track.  $ \mathbf{w_k} $ represents the process noise which allows for slack in the state dynamics, to account for when the motion of the UE differs from the constant velocity model. The UE logged $ \mathbf{P_k} $, i.e., the state covariance matrix (E$\left[\mathbf{x_k}\mathbf{x_k}^T\right]$), and $ \mathbf{x_k} $.

At sampling instant $ \mathbf{k} $, the EKF performs a \textit{prediction} step and an \textit{update} step. In the prediction step, the EKF predicts the state vector $ \mathbf{x_k} $ and covariance matrix $ \mathbf{P_k} $ of the UE from measurements up to sampling instant $ \mathbf{k-1} $, using \cite{Thacker_1998}
\begin{align}
	\widetilde{\mathbf{x_k}}&= \mathbf{F_{k-1}} \mathbf{x_{k-1}} + \mathbf{u_{k-1}} \label{eq:x_extrapolate}\\ 
	\widetilde{\mathbf{P_k}} &= \mathbf{F_{k-1}} \mathbf{P_{k-1}} \mathbf{F_{k-1}}^T + \mathbf{Q},\label{eq:P_extrapolate}
\end{align}
where $\mathbf{\widetilde{x_k}}$ is the predicted state vector, $ \mathbf{\widetilde{P_k}} $ is the predicted state covariance matrix, $ \mathbf{Q} $ is the process covariance matrix, equal to the covariance of the process noise $ \mathbf{w_k} $.

\begin{figure}
 	\centering
 	\includegraphics[width=0.30\textwidth]{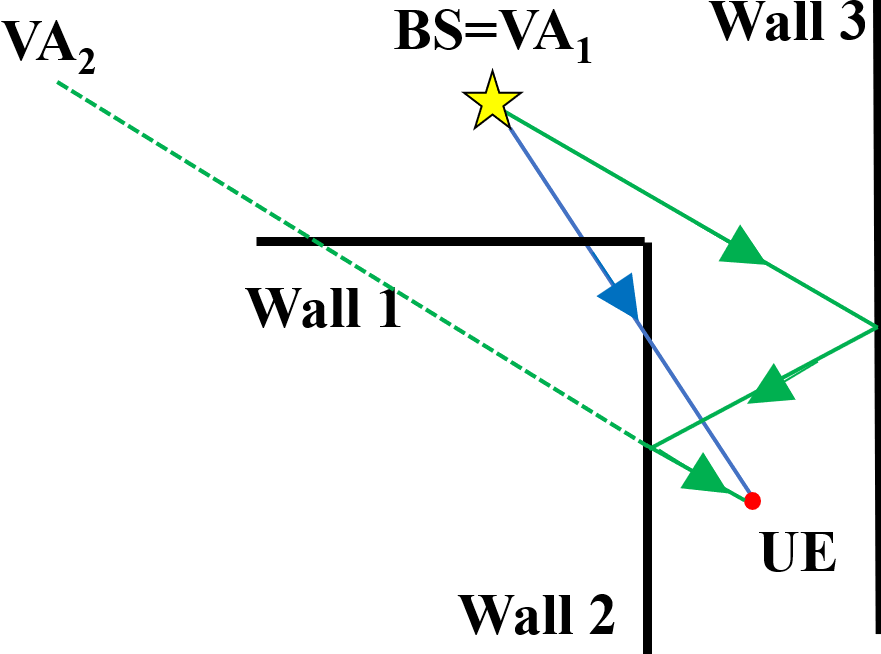}
 	\caption{ One virtual anchor for each MPC arriving at the UE is calculated, based on the path of the MPC predicted by MAP-AT.  }
 	\label{fig:virtual_anchors}
 \end{figure}

Using MAP-AT, the paths taken by each of the MPCs were predicted with the help of the measured ToF and AoA, using a site-specific map of the environment as seen in Fig \ref{fig:virtual_anchors} (by estimating the position of the UE via MAP-AT, as described in Section \ref{sec:MAP}). To implement an EKF, the geometric relationship between the current position of the UE and the measured ToF and AoA must be derived, for which the concept of virtual anchors (VA) was used. VAs are successive reflections of the BS on walls in the environment. The VA is treated as a LoS BS in place of the physical NLoS BS, as shown in Fig. \ref{fig:virtual_anchors}. Since the length of a MPC path does not change due to a reflection, the ToF of multipath that would have arrived at the UE from the VAs is equal to the ToF of the multipath arriving from the physical BS. Using VAs instead of physical BSs simplifies the geometric relationship between the current position of the UE and the measured ToF and AoA.
 
 The ToF and AoA measurements may be expressed in terms of the coordinates of the VA and the UE as follows: 

\begin{align}
\nonumber
	r &= \sqrt{(x_{VA}-x_k)^2+(y_{VA}-y_k)^2} \\\label{eq:r}
	&= c \times \text{ToF}\\
	\hat{n} &= \cos(\text{AoA}) = \dfrac{x_k-x_{VA}}{r}\label{eq:n},
\end{align}
where r is the 2D distance between the UE and VA, $(x_{VA},y_{VA})$ is the position of the VA, $ c $ is the speed of light, and $ \hat{n} $ is the cosine of the AoA of the MPC arriving from the VA. For mathematical convenience, $ r $ and $ \hat{n} $ are jointly represented by the measurement vector $\mathbf{z_k} = [ r\text{ } \hat{n} $]$ ^T $.

Once the prediction step is complete, in the update step, the EKF updates $\widetilde{\mathbf{x_k}}$ with measurements obtained at sampling instant $ \mathbf{k} $, to produce a final estimate of the UE state vector $ \widehat{\mathbf{x_k}} $. It can be shown that to minimize the least square error between $ \widehat{\mathbf{x_k}} $ and the true UE position and velocity \cite{Thacker_1998}, 
\begin{align}
 \widehat{\mathbf{x_k}}  = \widetilde{{\mathbf{x_k}}}+ \mathbf{K} \mathbf{i_k}. \label{eq:final_estimate}
\end{align}
 $ \mathbf{K} $ is the Kalman gain of the filter which will be defined shortly. The term $ \mathbf{i_k} $ is called the \textit{innovation} (also called \textit{measurement residual}) and is equal to the difference in $ \mathbf{z_k} $ (the measurement vector) and $ \widetilde{\mathbf{z_k}} $ (the value of the measurements calculated by replacing $ x_k $ and $ y_k $ with $ \widetilde{{{x_k}}} $ and $ \widetilde{{{y_k}}} $ in (\ref{eq:r}) and (\ref{eq:n})): 
 \begin{align}
 		\mathbf{i_k} = \mathbf{z_k} -\widetilde{\mathbf{z_k}}.
 \end{align}
 
As seen in (\ref{eq:final_estimate}), the final EKF estimate of the UE state vector, i.e., $ \widehat{\mathbf{x_k}} $, is a linear combination of measurement data collected up to time instant $ k-1 $ (in the form of $ \mathbf{\widetilde{x_k}} $) and new measurements collected at time instant $ k $. The weight of new measurements collected by the UE at time instant $ k $ in the linear combination is provided by $ \textbf{K} $ \cite{Thacker_1998}:
 \begin{align}
 \mathbf{K}=\mathbf{PH}^T (\mathbf{HPH}^T + \mathbf{R})^{-1} 	, \label{eq:Kalman_gain}
 \end{align}
where $ \mathbf{H} $ is the Jacobian of the measurements with respect to the extrapolated state vector $\widetilde{\mathbf{x_k}}$ (i.e. the multidimensional derivative of the measurements) and is given by:
\begin{align}
\nonumber
		&\mathbf{H}  = \begin{bmatrix} 
	\dfrac{d r}{d \widetilde{{x_k}}} & \dfrac{d r}{d \widetilde{{v_{xk}}}}	&\dfrac{d r}{d \widetilde{{y_k}}}  & \dfrac{d r}{d \widetilde{{v_{yk}}}} \\
		\dfrac{d \hat{n}}{d \widetilde{{x_k}}} & 	\dfrac{d \hat{n}}{d \widetilde{{v_{xk}}}}&	\dfrac{d \hat{n}}{d \widetilde{{y_k}}}&\dfrac{d \hat{n}}{d \widetilde{{v_{yk}}}}
	\end{bmatrix} 	\\
	&=\begin{bmatrix} 
	\dfrac{{{\widetilde{x_k}-x_{VA}}}} {r}& 0 &	\dfrac{{{\widetilde{y_k}-y_{VA}}}} {r} & 0\\
	\dfrac{(\widetilde{y_k}-y_{VA})^2}{r^3} & 0 & -\dfrac{(\widetilde{y_k}-y_{VA})(\widetilde{x_k}-x_{VA})} {r^3} & 0.
		\end{bmatrix} 	
\end{align}
 $ \mathbf{R} $, the measurement covariance matrix of ToF and AoA measurements, gives insight into the measurement noise levels.

\section{MAP-AT performance with Real-world Outdoor sub-THz measurements}\label{sec:results}
\begin{figure}[h]
	\centering
	\includegraphics[width=0.4\textwidth]{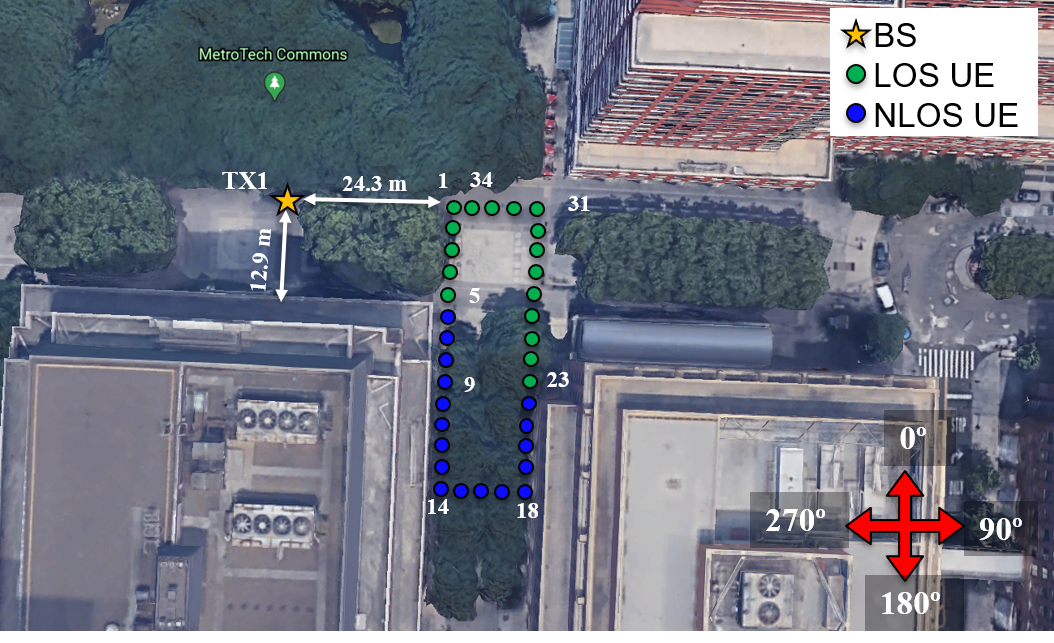}
	\caption{Map of the outdoor environment, depicting the outdoor LOS and NLOS locations where measurements were conducted at 140 GHz \cite{Xing21a}}
	\label{fig:outdoor_locations}
\end{figure}

\begin{figure}[h]
	\centering
	\includegraphics[width=0.25\textwidth]{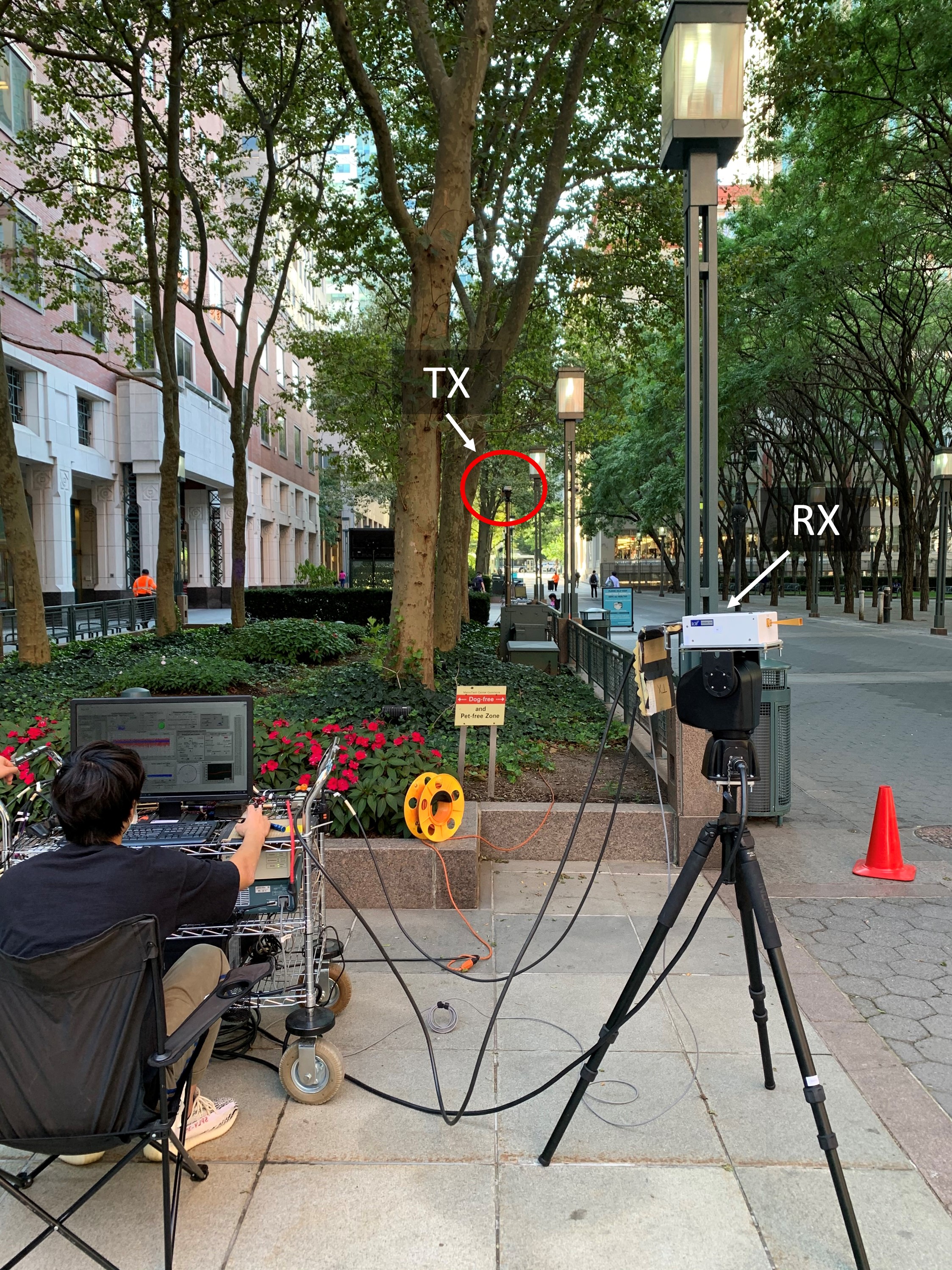}
	\caption{The LOS environment at UE location 1 is depicted in the image above.}
	\label{fig:L1RX1}
\end{figure}

\begin{figure}[h]
	\centering
	\includegraphics[width=0.35\textwidth]{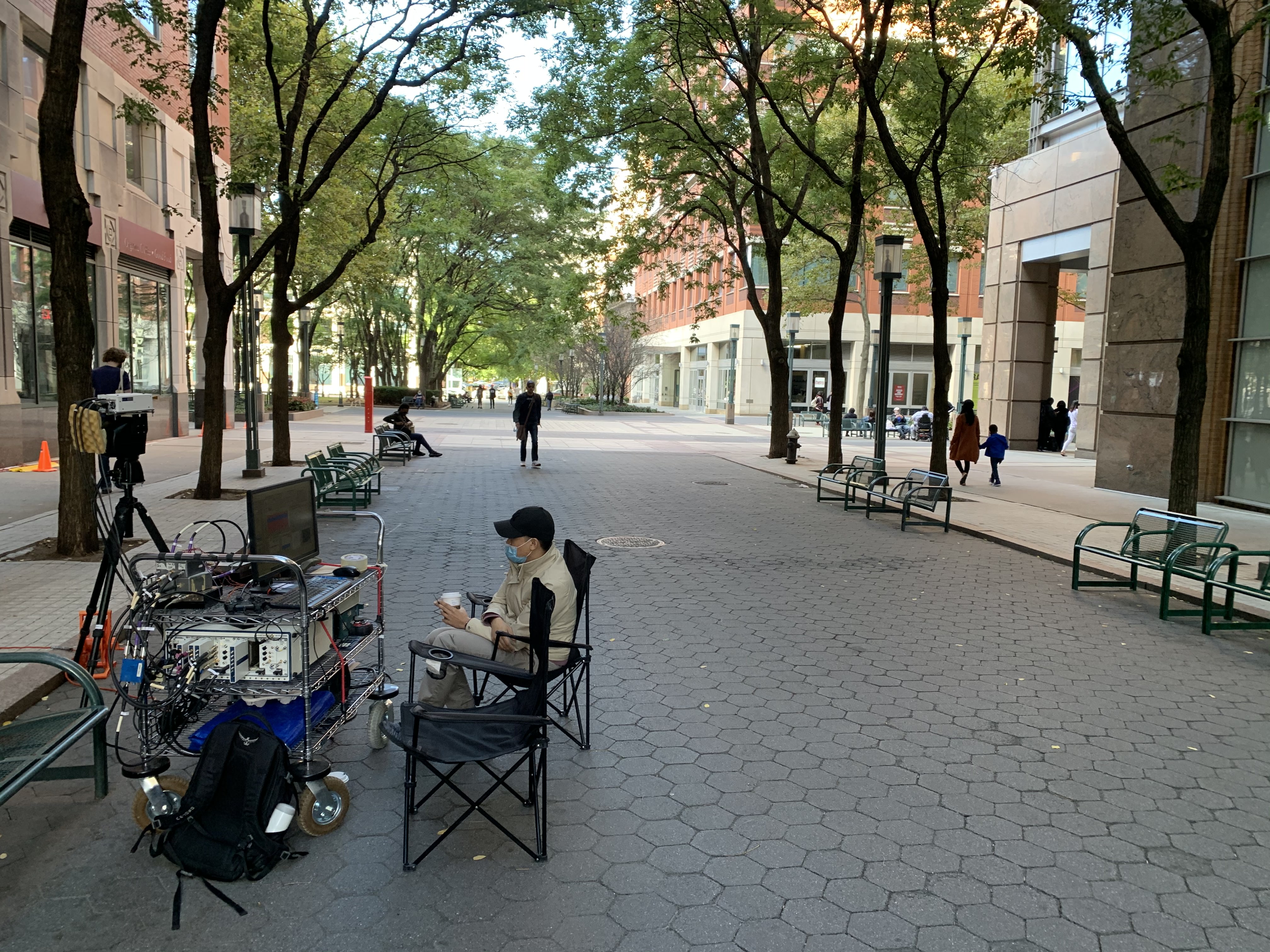}
	\caption{ UE location 15 NLOS with no direct signal path from the BS.}
	\label{fig:L1RX15}
\end{figure}

The localization performance of MAP-AT shall now be examined with real-world outdoor UMi measurements at 140 GHz. Measurements were conducted on the NYU Tandon engineering campus courtyard in downtown Brooklyn, New York as shown in Fig. \ref{fig:outdoor_locations}.  Out of the 34 UE locations in the 102 m long rectangular path, 17 locations were LOS, and 17 locations were NLOS. Fig. \ref{fig:L1RX1} depicts the environment of the LOS UE at location 1, while the typical environment experienced by a UE in NLOS is shown in Fig. \ref{fig:L1RX15}.

A wideband sliding correlator-based channel sounder was used to capture multipath PDPs and associated AoAs needed to implement MAP-AT, where the sounder had a null-to-null RF bandwidth of 1 GHz. Identical horn antennas at the BS and UE with 27 dBi gain and $ 8^\circ$ half-power beamwidths (HPBW) were used at 140 GHz, with a TX transmit power of 0 dBm, resulting in an effective isotropic radiated power (EIRP) of 27 dBm \cite{Xing_2018,Xing21a,Xing21b}. The horn antennas were mounted on electronically steerable gimbals with sub-degree accuracy in the azimuth and elevation plane. The TX was placed at a height of 4 m (the height of a lamppost), to replicate the location where cellular BSs could be deployed while the RX  was at a height of 1.5 m, the typical mobile UE height. Additional details about the measurement environment are provided in \cite{Xing21a}.

The measurements were conducted in a rectangular path of length 102 m, with a distance of 3 m between each measurement \cite{Xing21a,Ju21a}. Due to the bulky equipment at the RX, it was not possible to conduct measurements with the RX in motion, however, the static measurements were modeled as though a moving UE were sampling the wireless channel once every two seconds, assuming a walking speed of 1.5 m/s (since 1.5 m/s $ \times $ 2 seconds = 3 m). In a real deployment, UE velocity could be measured by doppler or by the sensors present in the mobile device such as the accelerometer and the gyroscope.

The channel sounder detected MPCs by conducting an exhaustive 3-D search at each of the 34 UE locations, by rotating the electronically controlled TX and RX gimbals \cite{Xing21a,Xing21b,Ju21a}. One to five MPCs were detected at each UE location except location 17, where no signal was received \cite{Ju21a}.

The channel sounder in \cite{Xing21a} measures relative timing of arriving MPCs via sliding correlation \cite{Xing_2018}. To calculate the absolute ToF of MPCs, ray tracing is required. The measured AoA was augmented with ToF predicted by NYURay, a 3-D mmWave ray tracer \cite{Kanhere19a}. Augmenting channel sounder measurements with ray tracing has proven valuable for producing statistical models as well as accurate site-specific models with absolute timing \cite{Samimi_2013,Samimi2015}. NYURay is calibrated to real-world mmWave measurements capable of providing accurate temporal, angular, and power measurements \cite{Kanhere19a}. Zero mean Gaussian noise with a standard deviation of 0.25 ns and $ 0.5 ^\circ $ was added to the measured ToF and AoA respectively to model measurement uncertainty. The augmented AoA and ToF measurements generated by NYURay were used by MAP-AT and the EKF for position location and tracking. The effect of increasing the standard deviation of ToF error to 0.5~ns on position location accuracy will be examined in Section \ref{sec:exp_results}.

\subsection{Extended Kalman Filter Parameter Selection}

Choosing appropriate values of the $ \mathbf{P_k} $, $ \mathbf{Q} $, and $ \mathbf{R} $ matrices described in Section \ref{sec:EKF} is necessary for optimal performance of the EKF and to ensure that the position location error is reasonably bounded \cite{Saho_2015}. 

The state covariance matrix $ \mathbf{P_k} $ in (\ref{eq:P_extrapolate}) is initialized based on the confidence in the initial state of the system. Since the initial UE position (location 1 in Fig. \ref{fig:outdoor_locations}) is a LOS location, the confidence in the initial position estimate of the UE is high and we set $ P_1 = 0.01  \mathbf{I_4}$, where $ \mathbf{I_4} $ is the 4$ \times $4 identify matrix. The initial position of the UE, $ x_1 $ was estimated via MAP-AT using the measured AoA and the ToF estimated by NYURay.
 
The process noise $  \mathbf{w_k} $ was modeled by a random acceleration, which perturbed the constant velocity assumed by the tracking model \cite{Saho_2015}. Thus the process noise is given by $ \mathbf{w_k} = (T^2/2 \quad T  \quad T^2/2 \quad T) \mathbf{w_a}$, where $ \mathbf{w_a}$ is a 2 $ \times $ 1 random vector containing the Gaussian random accelerations in the x and y directions with a mean of 0 and a standard deviation of $ \sigma_a= $0.05 m/s$ ^2 $. Assuming no cross-correlation between the random accelerations in the x and y directions (since the motion of the UE is independent in both directions), the covariance matrix of the random process is \cite{Saho_2015}: 
\begin{align}
\mathbf{Q}  &= \begin{bmatrix} 
T^4/4 & T^3/2 & 0 &0  \\
T^3/2 & T^2 &0 &0\\
0 & 0 & T^4/4 & T^3/2  \\
0 & 0 & T^3/2 & T^2 
\end{bmatrix} \sigma_a^2.
\end{align}

The measurement noise covariance matrix $ \mathbf{R} $ in (\ref{eq:Kalman_gain}) was set equal to a diagonal matrix, with entries equal to the variance of the ToF and AoA measurements since the errors in the ToF and AoA measurements were assumed to be uncorrelated \cite{Leibowicz_2000}. 
\subsection{Experimental Results} \label{sec:exp_results}
\begin{figure}
	\centering
	\includegraphics[width=0.35\textwidth]{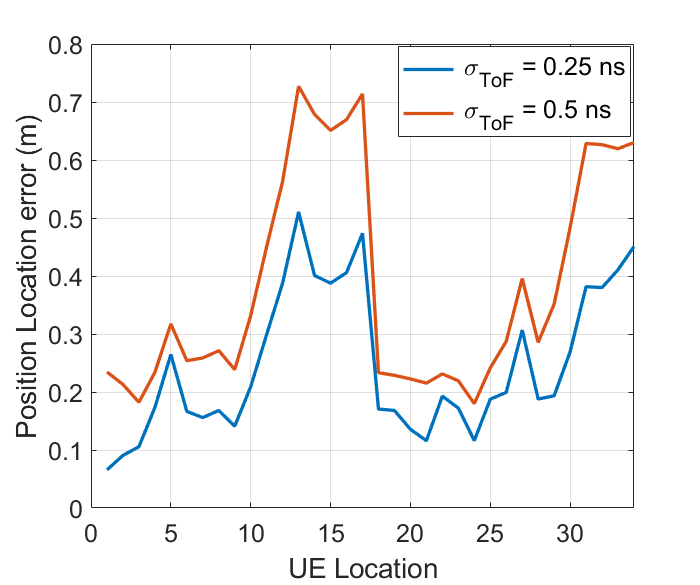}
	\caption{The variation of position location error over the 34 UE locations along the 102 m long rectangular track.}
	\label{fig:error}
\end{figure}

\begin{figure}
	\centering
	\includegraphics[width=0.4\textwidth]{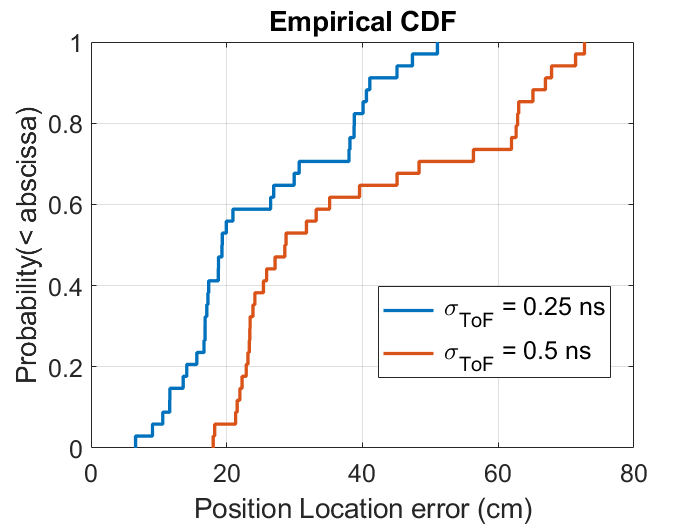}
	\caption{The CDF of the position location error along the rectangular track.}
	\label{fig:error_cdf}
\end{figure}

 Good localization results were obtained over the entire 102 m long rectangular track, with a mean error of 23.3 cm observed between the predicted and actual position location over the 17 LOS locations, and a mean error of 26.4 cm observed over the 17 NLOS locations at 140 GHz, with the TX-RX separation distance varying from 24.3 m to 52.8 m. Even though location 17 experienced an outage, the EKF was able to determine the location of the UE to an accuracy of 47.4 cm at location 17. A plot of the position location error at each UE location is provided in Fig. \ref{fig:error}, with a CDF of the position location errors provided in Fig. \ref{fig:error_cdf}. Increasing the standard deviation of the ToF error from 0.25 ns to 0.5 ns increased the mean localization error from 24.8 cm to 38.4 cm. It should be noted that these results are for a single BS, where additional base stations would likely improve the accuracy \cite{Kanhere19a}.

\subsection{Comparison of Accuracy With and Without Tracking}
Of the 34 UE locations along the rectangular path, the position of nine UE locations could be determined using MAP-AT alone, without using EKF, since two or more MPCs were received at the locations. Assuming a Gaussian ToF error with a standard deviation of 0.25 ns, the mean position location error over the nine UE locations was 7.39 cm using MAP-AT alone, and 10.39 cm using MAP-AT with EKF, with the TX-RX separation distance varying from 24.3 m to 46.8 m. 

\begin{figure}[h]
	\centering
	\includegraphics[width=0.4\textwidth]{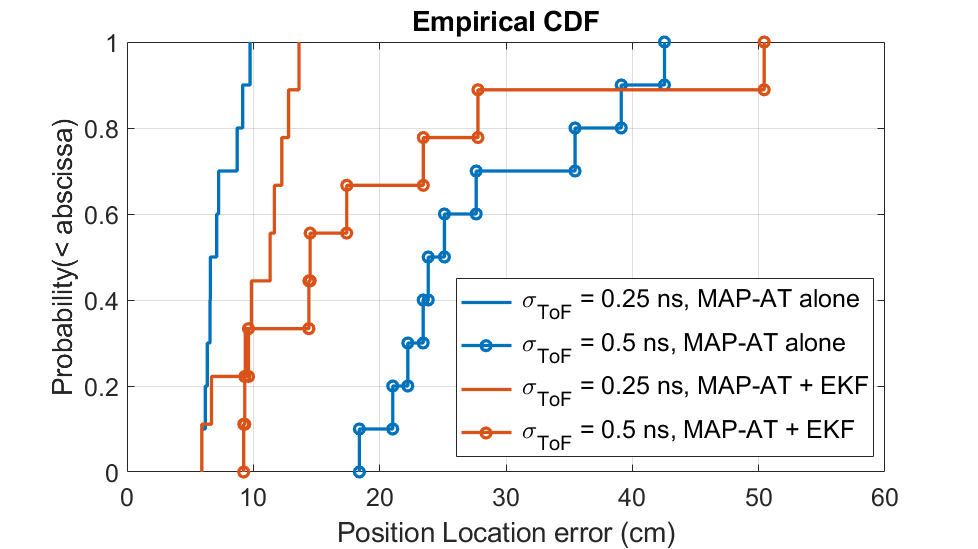}
	\caption{The UE position location error utilizing only MAP-AT vs. the accuracy with MAP-AT and the EKF. Although the mean position location error is similar, position location with EKF tracking has a greater variance in error.}
	\label{fig:with_vs_without_tracking}
\end{figure}

As seen in Fig. \ref{fig:with_vs_without_tracking}, UE tracking with the EKF did not significantly improve the localization accuracy of the nine UE locations which could be localized with MAP-AT alone, however, the EKF was critical for the 24 UE locations which received one MPC and UE location 17, which was in outage.
\section{Conclusion}\label{sec:conclusion}
MAP-AT determines the UE position in LOS and NLOS environments by combining the AoA and ToF of MPCs received at the UE using a site-specific map of the environment. This paper analyzed the performance of MAP-AT using real-world 140 GHz channel data. An EKF was used to track the position of the UE moving along a rectangular track, with only marginal improvement over not using EKF. The UE was in NLOS for half of the 34 measured locations since the direct signal from the BS was blocked by a building corner. A mean position location error of 24.8 cm was obtained at 140 GHz over 34 UE locations, with a TX-RX distance ranging from 24.3 m to 52.8 m. The EKF could predict the UE at a location of temporary outage (location 17) based on the tracked position and velocity of the UE. For the nine UE locations where two or more MPCs were received, the mean localization accuracy of MAP-AT without tracking (7.39 cm)  was similar to the error of MAP-AT with EKF tracking (10.39 cm).

Future work will analyze the improvement in position location accuracy by incorporating cooperative measurements, i.e. the measurement of signals transmitted between UEs.

\bibliographystyle{IEEEtran}
\bibliography{references}{}

\end{document}